\documentclass[
 reprint,
 amsmath,amssymb,
 superscriptaddress,
]{revtex4-1}

\usepackage{graphicx}
\usepackage{dcolumn}
\usepackage{bm}
\usepackage{lineno}

\begin{document}


\thanks{A footnote to the article title}%
\title{Second-harmonic focusing by nonlinear turbid medium via feedback-based wavefront shaping}

\author{Yanqi Qiao}
\affiliation{State Key Laboratory of Advanced Optical Communication Systems and Networks, Department of Physics and Astronomy, Shanghai Jiao Tong University, 800 Dongchuan Road, Shanghai 200240, China}
\affiliation{Key Laboratory for Laser plasmas (Ministry of Education), Collaborative Innovation Center of IFSA (CICIFSA), Shanghai Jiao Tong University, 800 Dongchuan Road, Shanghai 200240, China}

\author{Xianfeng Chen}
\email{xfchen@sjtu.edu.cn}
\affiliation{State Key Laboratory of Advanced Optical Communication Systems and Networks, Department of Physics and Astronomy, Shanghai Jiao Tong University, 800 Dongchuan Road, Shanghai 200240, China}
\affiliation{Key Laboratory for Laser plasmas (Ministry of Education), Collaborative Innovation Center of IFSA (CICIFSA), Shanghai Jiao Tong University, 800 Dongchuan Road, Shanghai 200240, China}

\author{Yajun Peng}
\affiliation{State Key Laboratory of Advanced Optical Communication Systems and Networks, Department of Physics and Astronomy, Shanghai Jiao Tong University, 800 Dongchuan Road, Shanghai 200240, China}
\affiliation{Key Laboratory for Laser plasmas (Ministry of Education), Collaborative Innovation Center of IFSA (CICIFSA), Shanghai Jiao Tong University, 800 Dongchuan Road, Shanghai 200240, China}

\author{Yuanlin Zheng}
\email{ylzheng@sjtu.edu.cn}
\affiliation{State Key Laboratory of Advanced Optical Communication Systems and Networks, Department of Physics and Astronomy, Shanghai Jiao Tong University, 800 Dongchuan Road, Shanghai 200240, China}
\affiliation{Key Laboratory for Laser plasmas (Ministry of Education), Collaborative Innovation Center of IFSA (CICIFSA), Shanghai Jiao Tong University, 800 Dongchuan Road, Shanghai 200240, China}

\date{\today}

\begin{abstract}

Scattering has usually be considered as detrimental for optical focusing or imaging. Recently, more and more research has shown that strongly scattering materials can be utilized to focus coherent light by controlling or shaping the incident light. Here, purposeful focusing of second-harmonic waves, which are generated and scattered from nonlinear turbid media via feedback-based wavefront shaping is presented. This work shows a flexible manipulation of both disordered linear and nonlinear scattering signals, indicating more controllable degrees of freedom for the description of turbid media. This technique also provides a possible way to an efficient transmission of nonlinear signal at a desired location in the form of a focal point or other patterns. With the combination of random nonlinear optics and wavefront shaping methods, more interesting applications are expected in the future, such as nonlinear transmission matrix, multi-frequency imaging and phase-matching-free nonlinear optics.

\end{abstract}

\maketitle

Optical focusing and imaging through complex media has attracted much interest for its applications in biological tissue microscopy, since a feedback-based wavefront shaping method was proposed in 2007 \cite{vellekoop2007focusing, vellekoop2008universal, vellekoop2015feedback, horstmeyer2015guidestar}. Great progress has been made for complex imaging, perfect focusing and even subwavelength focusing and imaging in the last decade \cite{conkey2012color, mosk2012controlling, vellekoop2010exploiting, park2013subwavelength, park2014full}. In fact, inelastic scattering light, especially nonlinear signal is also of great significance, such as focusing and imaging through turbid media via nonlinear feedback, as reported recently \cite{katz2011focusing, katz2014noninvasive, lai2015photoacoustically}. With a two-photon fluorescence feedback, spatiotemporal focusing and compression of chirped ultrashort pulses and a noninvasive imaging method were demonstrated. With a nonlinear photoacoustic signal feedback, an enhanced optical focusing on a blood layer through a scattering medium was demonstrated. However, it is noticeable that the turbid media applied in previous studies are usually linear materials, which means no frequency conversion occurred directly inside the medium. Here, we demonstrated the manipulation of second-order nonlinear signal (besides the linear part) generated and scattered from a turbid medium based on a wavefront shaping method, which indicates more controllable degrees of freedom (DOF) for the description of the turbid medium. Future focusing and imaging through turbid media may profit from this technique, with more information related to the scattering medium and the object.

Phase-matching (PM) or quasi-phase-matching (QPM) condition is required for high conversion efficiency in nonlinear optics, which may be hard to achieve in some cases \cite{boyd2008nonlinear}. In random nonlinear materials, a random QPM scheme was proposed \cite{baudrier2004random, fischer2006broadband, stivala2010random}, where the PM or QPM limitation can be loosened and the conversion efficiency could increase linearly with the interaction length $L$. When strong scattering was involved in the interaction, $L$ was much larger than the sample size and the emission signal would be totally divergent. For further investigation of the nonlinear optics inside the complex medium, a purposeful signal collecting way may be needed. Here, by manipulating the wavefront of the pump laser, we demonstrated an effective signal collecting way, with the divergent nonlinear signal focused at an assigned location in the form of a point or other complex patterns, such as two focuses and a focal line.

\begin{figure}[htb]
\centerline{
\includegraphics[width=8.0cm]{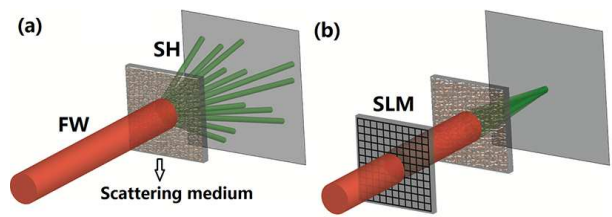}}
\caption{The concept of focusing SH signals generated and scattered from nonlinear medium via wavefront shaping. (a) Without shaping, the divergent SH signals form a disordered speckle pattern on the screen. (b) With an appropriate SLM phase mask applied before the scattering medium, an SH focal point can be generated on the screen.}
\end{figure}

\begin{figure*}[htb]
\centerline{
\includegraphics[width=15.5cm]{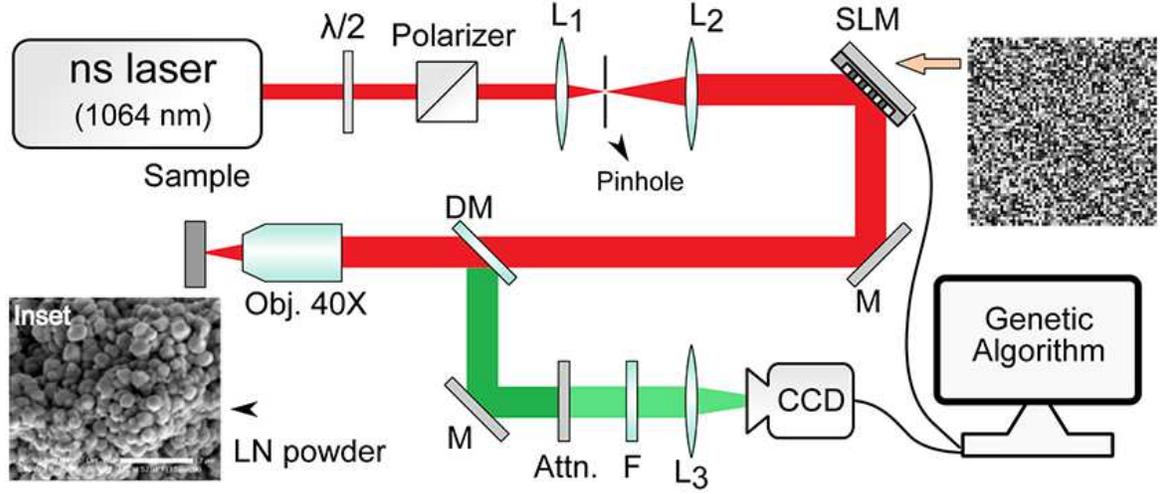}}
\caption{The experimental setup for divergent SH manipulation via wavefront shaping. $\lambda /2$: half-wave plate; $L_{1-3}$: lens, $f_{1-3}=50,200,100$ $mm$; M: mirror; DM: dichroic mirror; SLM: spatial light modulator; Attn.: attenuator; F: filter. Inset: A scanning electron microscopy image of the LN powder, deposited onto an ITO coated glass substrate (scale bar: 2 $\mu m$).}
\end{figure*}

In this Letter, the purposeful focusing of second-harmonic (SH) generated and scattered from superfine lithium niobate (LN) nanocrystal powder via feedback-based wavefront shaping method was demonstrated. The fundamental wave (FW) could also be focused using the same experimental setup via another round of optimization algorithm, as expected. That is to say, we could restore FW and SH focus respectively according to two masks applied on the spatial light modulator (SLM) after optimization. This technique provides more alternatives for future focusing and imaging applications. The SH focal spot in our experiment has a size of less than one pixel ($4.65^2$ $\mu m^2$) on the charge coupled device (CCD) camera, which means an accurate location and a high resolution. More complex manipulation of the divergent nonlinear signal was also demonstrated in the same experimental setup, such as two SH focuses and an SH focal line.

The concept of SH focusing via feedback-based wavefront shaping is shown in Fig. 1. Firstly, without FW wavefront shaping, the SH signals generated and scattered from the nonlinear turbid medium were divergent and manifested themselves as disorganized speckles on the screen [Fig. 1(a)]. Then an SLM shown in Fig. 1(b) was applied to shape the FW. When an optimization algorithm was applied, a corresponding mask would be found to form an SH focus. Obviously, there was also FW scattered from the medium and another phase mask could be found to make FW focusing, which has been demonstrated in some previous experiments \cite{vellekoop2015feedback}. More complex manipulation as mentioned above was achieved with different optimization algorithms and masks.

The experimental setup is illustrated in Fig. 2. The light source was a high energy diode pumped all-solid-state Q-switched laser at the wavelength of 1064 $nm$ (10 $ns$, 1 $kHz$). A half-wave plate and a Glan-Taylor polarizer were used for polarization and power control. The SLM had a resolution of $512\times512$ pixels, each with a rectangular area of $19.5\times19.5$ $\mu m^2$. The FW was spatially filtered and expanded to fit the SLM aperture. Then the shaped wave was focused onto the sample by an objective ($40\times$, $NA=0.65$) after a dichroic mirror. The generated and scattered SH signal was collected in the reflected direction and imaged on a multi-spectral 2-channel CCD camera (JAI, AD-080 GE) by a lens after passing an attenuator and a filter. The CCD camera had a resolution of $1024\times768$ pixels, each with a rectangular area of $4.65^2$ $\mu m^2$.

Here, LN nanocrystal powder was used as the second-order nonlinear scattering medium, which was prepared by the solid state reaction method using niobium pentoxide ($Nb_2O_5$) and lithium acetate dihydrate ($C_2H_3O_2Li\cdot2H_2O$) as reactants \cite{su2010preparation}. Then the electrophoresis method was applied to deposit the powder onto an indium-tin oxide (ITO) coated glass substrate \cite{jeon1996electrophoretic}. The LN particles had a maximum size of $\sim$400 $nm$ (see Fig. 2 inset) and the powder layer had a dimension of around $30\times30\times0.1$ $mm^3$. The scattering mean free path of the LN powder layer was estimated to be $l\cong1.1\lambda$ from a coherent backscattering experiment \cite{van1985observation, wolf1985weak, de1996coherent}.

The SH speckle pattern detected by the CCD camera was recorded and served as a feedback to optimize the SLM phase mask. A genetic algorithm was applied to realize the optimization, for its robustness to the noisy environments \cite{conkey2012genetic}. During the optimization, $512\times512$ pixels on the SLM were divided into $64\times64$ segments to shape the FW wavefront. To start with, 100 masks were generated randomly as the first generation. Then they ranked according to the SH intensity at the target location. With a higher rank, the mask would has a higher chance to be selected as the parent to create the next generation, containing another 100 masks. There were some mutation during the creation, which means several segments of the mask would mutate in the next generation. Next, the genetic process was repeated and the latter generation would generally do better than the former, which indicated an optimization process or a manipulation of the divergent SH signal. Finally, a focal point or other designed patterns was formed.

\begin{figure}[htb]
\centerline{
\includegraphics[width=8.0cm]{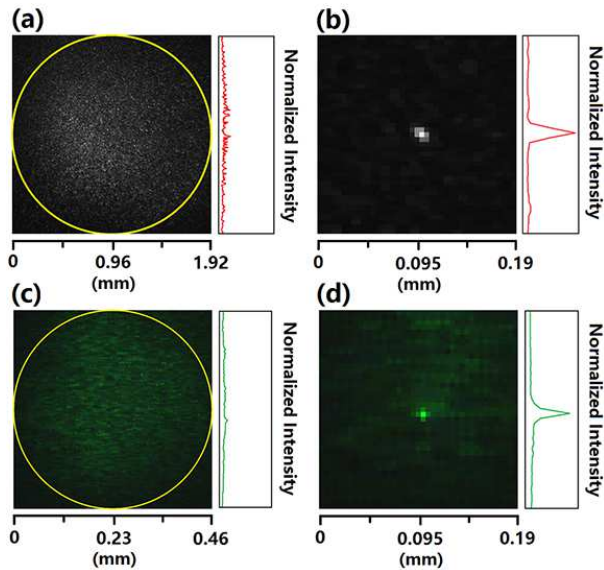}}
\caption{(a, c) Before optimization, the FW and SH speckle pattern has a size of around 2.9 $mm^2$ and 0.65 $mm^2$, respectively. (b, d) After optimization via a genetic algorithm, the FW and SH speckle pattern turns into a focal point. The SH focus size is equivalent to that of only one pixel size (21.6 $\mu m^2$) on the CCD screen.}
\end{figure}

Before manipulating the SH signals, we switched the CCD camera to infrared channel and used a long-pass filter to check the system. Although the FW reflected from the dichroic mirror was very weak, it was still intense enough for CCD detection. After several generations, an FW focus was observed on the screen. Figure 3(a) and 3(b) show the detected FW patterns before and after optimization, respectively. The yellow circle marked the original speckle size to be around 2.9 $mm^2$. The optimized focus is around several pixels size (each with a rectangular area of 21.6 $\mu m^2$). This process was fast and repeatable, which confirmed the stability of our system. When considering the SH signals, we switched the CCD camera to visible channel and used a short-pass filter as illustrated in Fig. 2. The speckle size before optimization was around 0.65 $mm^2$ as shown in Fig. 3(c), and after optimization, the focus size was only one pixel size as shown in Fig. 3(d). The result indicated an accurate location and a high resolution of SH focus which is desirable in further applications. It is worth mentioning that the focuses could be restored after a long time with the applying of the masks achieved via optimization.

The enhancement factor $\eta$ of the SH intensity at the target location was also detected as the generation number increased. As expected, $\eta$ increased generally with the generation number until saturation, which was usually a local maximum value and determined by the segment number, the particle size and the algorithm details \cite{conkey2012genetic}. A typical enhancement in our SH experiments was estimated to be 27 after 300 generations as shown in Fig. 4. In fact, a continuous sequential algorithm may do better in the enhancement at the cost of time spent and vulnerability to environmental noise \cite{conkey2012genetic}. Other algorithms, such as partitioning and transmission matrix, should also be effective in the optimization for different applications \cite{conkey2012genetic, vellekoop2008phase}.

Some typical results of more complex manipulation of the scattered SH signals were shown in Fig. 5. Firstly, the average intensity of two assigned locations served as the feedback for optimization. After around 600 generations, two SH focuses were observed on the CCD screen, as shown in Fig. 5(a) and 5(c). The point-to-point distance is 93 $\mu m$ (20 pixels) and the enhancement factor $\eta$ is around 15. Similarly, if the average intensity of a series of pixels was set as the feedback, we would find a focal line with a length of about 51 $\mu m$ (11 pixels) on the screen, as shown in Fig. 5(b) and 5(d). In fact, the more complex manipulation we hope, the more time (generations), the smaller segments and scattering particles were needed. It is expected that any complex manipulation was theoretically possible. The flexible manipulation of the SH signals also indicates a simpler way to generate nearly arbitrary designed nonlinear patterns compared to no-feedback-based wavefront shaping and domain design methods \cite{qin2008wave, Ellenbogen2009Nonlinear, shapira2012two, hong2014nonlinear, liu2016nonlinear, liu2016scattering}.

\begin{figure}[htb]
\centerline{
\includegraphics[width=7.5cm]{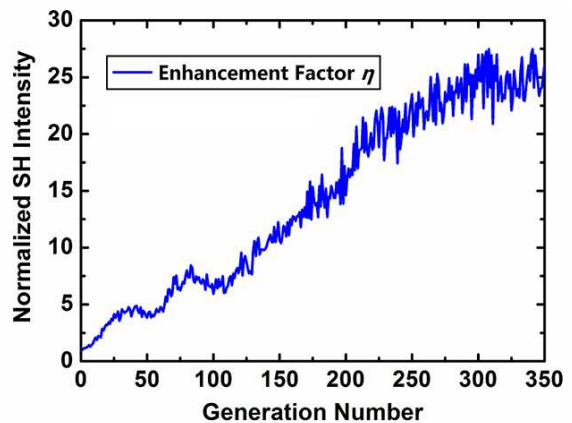}}
\caption{The enhancement factor $\eta$ of the normalized SH intensity at the target location versus the generation number. A typical value was estimated to be 27 after 300 generations.}
\end{figure}

In our experimental design, LN nanoparticle powder was chosen for its large second-order susceptibility and high refractive index ($n\approx2.2$) for strong scattering \cite{boyd2008nonlinear}. The maximum size of the particle was around 400 $nm$, which is smaller than the coherent interaction length $L_c$ in LN crystal ($L_c$ is about several microns). This fact ensures that every single particle serves as an SH source and generates maximum emission signal. Besides, the smaller the particle size is, the more manipulation DOF is available. Other nonlinear scattering materials are believed applicative in the same experiments. Nonlinear processes, such as third-harmonic generation (THG), four-wave mixing (FWM), and coherent anti-Stokes Raman scattering (CARS) are also expected to behave similarly with proper configurations \cite{boyd2008nonlinear}. However, a main limitation on future applications may be the optimization speed when manipulating complex patterns. For example, it took $\sim$10 minutes of optimization to get the SH focal line in our experiment. This problem can be solved with faster software-hardware configuration and better algorithms in the future.

\begin{figure}[htb]
\centerline{
\includegraphics[width=8.0cm]{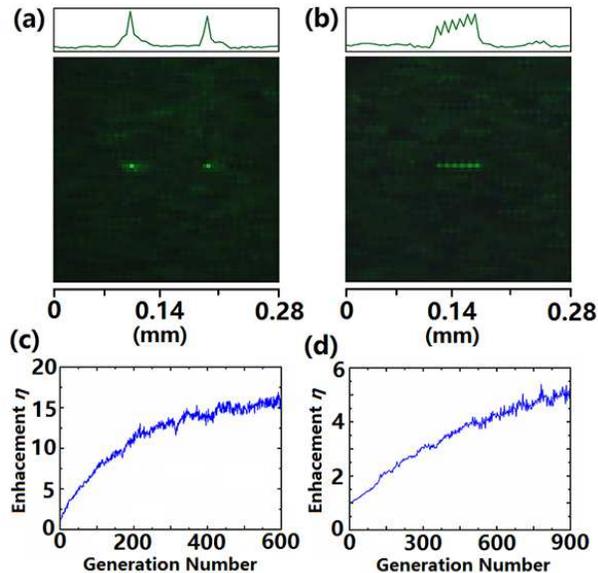}}
\caption{(a) Two SH focuses obtained on the CCD screen after approximate 600 generations. (b) A focal line formed on the screen after optimization, with a length about 51 $\mu m$ (11 pixels). (c, d) The enhancement factors of the two focuses ($\eta\approx15$) and the focal line ($\eta\approx5$), respectively.}
\end{figure}

In conclusion, we have reported the purposeful focusing of SH generated and scattered from superfine LN nanocrystal powder via feedback-based wavefront shaping method. A genetic algorithm was applied in the optimization for its robustness. The FW and SH focuses could be restored, respectively, according to the masks applied on the SLM by optimization. It is worth mentioning that the SH focus was only one pixel size ($4.65^2$ $\mu m^2$) which indicates an accurate location and a high resolution. The enhancement factor $\eta$ of the SH intensity at the target location was measured to be around 27 after 300 generations. Besides, more complex manipulation of the scattered SH signals were demonstrated in the experiment, such as two SH focuses and a focal line. Other nonlinear scattering materials or nonlinear processes are believed to behave similarly under proper configuration.

This work indicates more controllable DOF for the description of turbid media. In fact, the transmission matrix (TM) method was very popular in describing the scattering properties of linear media \cite{beenakker1997random, popoff2010measuring}. Image recovery through scattering media becomes possible and easy based on this method, except that the sample needs to be very stable in a certain time interval \cite{popoff2010image, popoff2011controlling}. With nonlinear scattering media, a nonlinear TM may be expected and future focusing and imaging through turbid media may profit from this expansion. A point spread function (PSF) method is also effective on describing scattering media for a high speed color imaging besides TM method, which may be applied in the nonlinear medium and nonlinear imaging \cite{zhuang2016high}. This technique also provides a possible way to transmit efficient nonlinear signal at a desired location in the form of a point or other patterns. With this combination of random nonlinear optics and wavefront shaping method, more and more interesting applications can be envisioned in the future, such as multi-frequency imaging and phase-matching-free or domain-design-free nonlinear optics.\\

This work is supported in part by the National Natural Science Foundation of China under Grant Nos. 61235009 and 11421064.


%

\end{document}